\documentclass[12pt]{iopart}
\usepackage{iopams}
\usepackage{cite}

\begin{document}
\title[On ... the uncertainty relations on circle]{On a subject of diverse
improvisations: \\ The uncertainty relations on a circle}
\author{S. Dumitru}
\address{Department  of Physics, ``Transilvania'' University,
B-dul Eroilor 29, R-2200 Brasov, Romania}
\ead{s.dumitru@unitbv.ro}
\begin{abstract}
The disputed question of uncertainty relations (UR) on a circle is
regarded as a particular element of a more general problem which
refers to the quantum description of angular observables $L_z$ and
$\varphi$. The improvised $L_z-\varphi$ UR are found to be affected by
unsourmontable shortcomings. Also in contradiction with a largely
accepted belief it is proved that the usual procedures of quantum
mechanics are accurately applicable for the $L_z-\varphi$ pair.
The applicability regards both the known circular motions and the
less known non-circular rotational motions. The presented facts
contribute as arguments to the following indubitable conclusions:
(i) the traditional interpretation of UR must be denied as an
incorrect doctrine, (ii) for a natural physical consideration of
the the $L_z-\varphi$ pair the results from  the usual quantum
mechanics are sufficient while the improvised $L_z-\varphi$ UR
must be rejected as senseless formulas and (iii) the descriptions
of quantum measurements have to be done in a framework which is
distinct and additional in respect with the usual quantum
mechanics.
\end{abstract}
\submitto{JPA}
\maketitle

\section{Introduction}
A recent work \cite{1} reminds the fact that, in spite of its
importance and long history, the problem of uncertainty relations
(UR) on a circle still remains open. Then, by using an ingenious
labour, it is improvised a new such UR which, in its essence
regards the angular variables $L_z$ (or $J$) and $\varphi$
($z$-component of angular momentum and azimuthal angle). So the
known class \cite{2,3,4,5,6,7,8,9,10,11,12,13,14,15,16,17} of
improvised $L_z-\varphi$ UR is enlarged. The relations from the
mentioned class are introduced through esoteric and disonant
considerations. The respective considerations seem to be motivated
exclusively by the preocupation of obtaining an accordance with
the traditional interpretation of UR (TIUR). They are also
associated with the belief that the usual quantum mechanics (QM)
procedures (i.e operator-based and Fourier transform approaches)
are not applicable for the $L_z-\varphi$ pair.

In this paper, in Sec.II, we point out the fact that the class of
improvised $L_z-\varphi$ UR are affected by shortcomings
(dissimilarities and deficiencies). The respective shortcomings
are unsurmountable because they cannot be avoided through valid
arguments (of physical and/or mathematical nature).

Contrary to the above-alluded belief, in Sec.III, we reveal that
the usual QM procedures are applicable directly, on just ways, to
the $L_z-\varphi$ pair. We show that the respective procedures
work naturally (i.e. without any esoteric considerations) for all
the rotational motions which the improvised $L_z-\varphi$ UR refer
to. Moreover we find that the same procedures are also valid for
the cases of non-circular rotations, never refered by the
mentioned impovisations. Those cases regard the quantum torsion
pendulum (QTP) and degenerate spatial rotations.

For QM problems the findings from Sec.III are things of technical
nature - i.e. they report bare mathematical results. But as it is
known QM is also involved in controversial questions connected
with TIUR. The significance of the mentioned findings for TIUR and
related questions are dicused in Sec.IV. The respective dicussions
are focused on the known fact that the disputes regarding the
$L_z-\varphi$ pair originate from TIUR and from the associated
belief about the quantum measurements.  Such a focusing dicloses a
lot of incontestable arguments which, taken together with other
doubtless facts \cite{18,19,20,21,22,23,24,25,26}, entail the
conclusion that TIUR must be denied as an incorrect doctrine.
Associated with the mentioned conclusion we find that the usual QM
procedures are sufficient for a natural and correct description of
the observables $L_z$ and $\varphi$. Consequently the improvised
$L_z-\varphi$ UR have to be rejected as formulas without any
physical significance. In  the same context we opine that the
descriptions of the quantum measurements necesitate a distinct
framework, additional to the usual QM.

\section{Improvised $L_z-\varphi$ relations and related shortcomings}
The largely known searches and disputes about the $L_z-\varphi$ UR
refer exclusively to the restricted class of circular motions. As
concrete  examples (usually not listed explicitely in the
literature) the mentioned class includes: the motion of a particle
on a circle, the plane rotations of a rotator with fixed axis and
the completely indexed spatial rotations. The alluded kind of
spatial rotations refer to: (i) a particle on a sphere, (ii) a
rotator with mobile axis and (iii) an electron in a hidrogen-like
atom. By completely indexed we denote a situation for which the
corresponding wave function has unique values for all the implied
quantum numbers (i.e. for $m$ and $l$ in in the cases (i)  and
(ii) respectively for $m$, $l$ and $n$ in the case (iii) - $m$,
$l$ and $n$ being the magnetic, orbital and principal quantum
numbers).

For all circular motions the part of the wave function regarding
the $L_z-\varphi$ pair has the form
\begin{equation}\label{eq:1}
  \Psi_m(\varphi)=\frac{1}{\sqrt{2\pi}}\, e^{im\varphi} \; ,
  \qquad \varphi\in [0,2\pi]
\end{equation}
with a single value for the integer number $m$. On the other hand,
according to the usual QM procedures, the observables $L_z$ and
$\varphi$ are described by the operators
\begin{equation}\label{eq:2}
  \hat{L}_z=-i\hbar \frac{\partial}{\partial\varphi} \; ,
  \qquad \hat{\varphi}=\varphi\,\cdot
\end{equation}
respectively by the commutation relation
\begin{equation}\label{eq:3}
  [ \hat{L}_z,\hat\varphi ] = -i\hbar
\end{equation}
For the observables $L_z$ and $\varphi$ in the situations
described by (\ref{eq:1}) the corresponding standard deviations
$\Delta L_z$ and $\Delta\varphi$ (defined  in the usual sense -
see also the next section) have the expressions
\begin{equation}
  \label{eq:4}
  \Delta L_z=0 \; , \qquad \Delta\varphi = \frac{\pi}{\sqrt{3}}
\end{equation}
But such expressions are incompatible with the formula
\begin{equation}
  \label{eq:5}
  \Delta L_z \cdot \Delta\varphi \geqslant \frac{\hbar}{2}
\end{equation}
required by TIUR for two observables described by conjugated operators as
the ones given
by (\ref{eq:2}) and (\ref{eq:3}).

In order to avoid the mentioned incompatibility many scientists
sustained the belief that for the $L_z-\varphi$ pair the usual QM
procedures do not work correctly. Consequently it was accredited
the idea that formula (\ref{eq:5}) must be prohibited and replaced
with other $L_z-\varphi$ UR. So, along the years, instead of
(\ref{eq:5}) there were promoted
\cite{1,2,3,4,5,6,7,8,9,10,11,12,13,14,15,16,17} a lot of
improvised $L_z-\varphi$ UR such are
\begin{equation}
  \label{eq:6}
  \frac{\Delta_z L_z\cdot\Delta\varphi}{1-3(\Delta\varphi/\pi)^2}
  \geqslant 0.16\hbar
\end{equation}
\begin{equation}
  \label{eq:7}
  \frac{(\Delta L_z)^2\cdot (\Delta\varphi)^2}{1-(\Delta\varphi)^2}
  \geqslant \frac{\hbar^2}{4}
\end{equation}
\begin{equation}
  \label{eq:8}
  (\Delta L_z)^2+\left(\frac{\hbar}{2}\alpha\right)^2\cdot
  (\Delta\varphi)^2 \geqslant
  \frac{\hbar^2}{2}\left[\left(\frac{9}{\pi^2}+\alpha^2\right)^{1/2}-
  \frac{3}{\pi^2}\right]
\end{equation}
\begin{equation}
  \label{eq:9}
  \frac{\Delta_z L_z\cdot\Delta\varphi}{1-3(\Delta\varphi/\pi)^2}\geqslant
  \hbar\, \frac{2}{3}\left(\frac{V_{\min}}{V_{\max}}\right)
\end{equation}
\begin{equation}
  \label{eq:10}
  (\Delta L_z)^2\cdot (\Delta\sin\varphi)^2\geqslant \frac{\hbar^2}{4}
  \langle\cos^2\varphi\rangle
\end{equation}
\begin{equation}
  \label{eq:11}
  (\Delta L_z)^2\cdot (\Delta\cos\varphi)^2\geqslant \frac{\hbar^2}{4}
  \langle\sin^2\varphi\rangle
\end{equation}
\begin{equation}
  \label{eq:12}
  \Delta L_z \cdot \Delta\chi \geqslant \frac{\hbar}{2}
\end{equation}
\begin{equation}
  \label{eq:13}
  \Delta L_z\cdot \Delta\varphi \geqslant
  \frac{\hbar}{2}\left\vert\langle\varepsilon(\varphi)\rangle\right\vert
\end{equation}
\begin{equation}
  \label{eq:14}
  (\Delta L_z)^2+\hbar^2(\Delta\varphi)^2\geqslant \hbar^2
\end{equation}
\begin{equation}
  \label{eq:15}
  \Delta L_z \cdot \Delta\varphi \geqslant \frac{\hbar}{2} \left\vert 1-2\pi
  \vert\Psi(2\pi)\vert^2\right\vert
\end{equation}
In (\ref{eq:8}) $\alpha$ is a real parameter. In (\ref{eq:9})
$V_{\min}$ and $V_{\max}$ represent the minimum respectively the
maximum values of $V(\beta)=\int_{-\pi}^{\pi}\beta
\vert\Psi(\alpha+ \beta)\vert^2 d\varphi$ where
$\beta\in[-\pi,\pi]$ and $\Psi$ denotes the wave  function. In
(\ref{eq:12}) $\chi=\varphi+2\pi N$,
$\Delta\chi=\left[2\pi^2\left(\frac{1}{12}+N^2-N_1^2+
N-N_1\right)\right]^{1/2}$, while $N$ and $N_1$ with $N\neq N_1$,
denote  two arbitrary integer numbers. In (\ref{eq:13})
$\varepsilon(\varphi)$ is a complicated expression of $\varphi$.
Relation (\ref{eq:14}) is written in original version \cite{1}
with $\hbar=1$ and $L_z=J$.

A minute examination of the facts shows that, in its essence, the
set of improvised relations (\ref{eq:6})-(\ref{eq:15}) is affected
by the following shortcomings (\textbf{Shc}):
\begin{itemize}
\item \textbf{Shc.1}: None of the relations  (\ref{eq:6})-(\ref{eq:15})
is unanimously accepted as a correct version for theoretical
$L_z-\varphi$ UR.
\item \textbf{Shc.2}: From a mathematical perspective the relations
(\ref{eq:6})-(\ref{eq:15}) are not mutually equivalent.
\item \textbf{Shc.3}: The relations (\ref{eq:6})-(\ref{eq:14}) do not have
correct supports in the usual formalisn of QM (that however works
very well in a huge number of applications).
\item \textbf{Shc.4}: In fact the considerations implied in the promotion of
the relations (\ref{eq:6})-(\ref{eq:14}) do not have real physical
motivations (argumentations).
\end{itemize}

\emph{Observation} We do not associate the formula (\ref{eq:15})
with \textbf{Shc.3} and \textbf{Shc.4} because it proves itself to
be a relation derivable from the usual QM procedures (see the
relation (\ref{eq:52}) in Sec.III).

Relations (\ref{eq:6})-(\ref{eq:15}) refer to the circular motions
in which $L_z$ and $\varphi$ are in postures of basic observables.
But $L_z$ and $\varphi$ are also in similar postures in the cases
of non-circular rotations (NCR). Within the class of NCR we
include: motion of quantum torsion pendulum (QTP) and degenerated
(or incompletely indexed) spatial rotations. The alluded
degenerate rotations refer to: (i) a particle on a sphere, (ii) a
rotator with mobile axis and (iii) an electron in a hydrogen-like
atom. By degenerate motions we refer to the situations when the
energy of sistem is well precised while the non-energetic quantum
numbers take all the permited values. Such numbers are $m$ in the
cases (i) and (ii) respectively $l$ and $m$ in the case (iii).

From the class of NCR let us firstly refer to the motions of a QTP, which
\cite{20,23,26} is  a quantum harmonic oscilator described by the
Hamiltonian
\begin{equation}
  \label{eq:16}
  \hat{H}=\frac{1}{2I}\,\hat{L}_z^2+\frac{I\omega^2}{2}\,\varphi^2=
  -\frac{\hbar^2}{2I}\frac{\partial^2}{\partial\varphi^2} +
  \frac{I\omega^2}{2}\,\varphi^2\cdot
\end{equation}
with $I$ = moment of inertia and $\omega$ = angular frequency.
Consequently the states of QTP are described by the wave functions
\begin{equation}
  \label{eq:17}
  \Psi_n(\varphi)=\Psi_n(\xi)\sim \exp \left\{-\frac{\xi^2}{2}\right\}
  \mathcal{H}_n(\xi)\; ,
  \qquad \xi=\varphi\sqrt{\frac{I\omega}{\hbar}}\in(-\infty,\infty)
\end{equation}
and energies
\begin{equation}
  \label{eq:18}
  E_n=\hbar\omega\left(n+\frac{1}{2}\right)
\end{equation}
In (\ref{eq:17}) and (\ref{eq:18}) $n=0,1,2,\ldots$ = oscilation
quantum number and $\mathcal{H}_n(\xi)$ denote the Hermite
polinomials. Also in the case of QTP the observables $L_z$ and
$\varphi$ are described by the operators presented in (\ref{eq:2})
and (\ref{eq:3}). Then, by  using (\ref{eq:17}), similarly with
the case of recti-linear oscilator \cite{27,23} for the standard
deviations $\Delta L_z$ and $\Delta\varphi$ one obtains the
expressions
\begin{equation}
  \label{eq:19}
  \Delta L_z=\sqrt{\hbar I\omega\left(n+\frac{1}{2}\right)}\; , \qquad
  \Delta\varphi=\sqrt{\frac{\hbar}{I\omega}\left(n+\frac{1}{2}\right)}
\end{equation}
With these expressions one finds that for QTP the $L_z-\varphi$
pair satisfies the ``prohibited'' formula (\ref{eq:5}).

From the same class of NCR now let us discuss the cases of
degenerate motions such one finds \cite{26} for a particle on a
sphere as well as for a rotator with mobile axis (the problem of
an atomic electron can be discussed in a completely similar way).
In both the mentioned cases a degenerate state corresponds to an
energy
\begin{equation}
  \label{eq:20}
  E_l=\frac{\hbar^2}{2I}\, l(l+1)
\end{equation}
where $I$ is the moment of inertia and $l$ denotes  the orbital
quantum number, which has a well-precised value. Such a state is
degenerate in respect with the magnetic quantum number $m$ that
has  the pemited values $m=0,\pm 1,\pm 2,\ldots,\pm l$.
Consequently the wave function of the respective state is of the
form
\begin{equation}
  \label{eq:21}
  \Psi_l(\theta,\varphi)=\sum_{m=-l}^{l}c_m\, Y_{lm}(\theta,\varphi)\; ,
  \qquad \varphi\in [0,2\pi]
\end{equation}
where $Y_{lm}(\theta,\varphi)$ are the spherical functions and
$c_m$ denote complex coefficients, which satisfy the condition
\begin{equation}
  \label{eq:22}
  \sum_{m=-l}^{l}\vert c_m \vert^2=1
\end{equation}
By using the wave functions $\Psi_l(\theta,\varphi)$ given by
(\ref{eq:21}) respectively the operators $\hat{L}_z$ and
$\hat{\varphi}$ as defined in (\ref{eq:2})-(\ref{eq:3}) one finds
\begin{equation}
  \label{eq:23}
  (\Delta L_z')^2=\sum_{m=-l}^{l}\vert c_m \vert^2 \hbar^2 m^2 -
  \left[\sum_{m=-l}^{l}\vert c_m \vert^2 \hbar m\right]^2
\end{equation}
\begin{equation}
  \label{eq:24}
  \fl (\Delta\varphi)^2=\sum_{m=-l}^{l}\sum_{m'=-l}^{l} c_m^* c_{m'}
  (Y_{lm},\hat\varphi^2\, Y_{lm'})-\left[\sum_{m=-l}^{l}\sum_{m'=-l}^{l}
  c_m^*c_{m'} (Y_{lm},\hat\varphi Y_{lm'})\right]^2
\end{equation}
Where $(f,g)$ denote the scalar product of the functions $f$ and
$g$ (for notations see also the next section). With (\ref{eq:23})
and (\ref{eq:24}) in the cases described by (\ref{eq:21}), one
finds that it is possible for the ``prohibited'' relation to be
verified. The respective possibility is conditioned by the
concrete values of the coefficients $c_m$.

The above presented facts in connection with NCR, regarded
together with the mentioned discussions about the improvised
$L_z-\varphi$ UR (\ref{eq:6})-(\ref{eq:15}), induce the following
questions (\textbf{Q}):
\begin{itemize}
\item \textbf{Q.1}: What is the significance of the respective facts for
TIUR?
\item \textbf{Q.2}: Are the usual QM procedures really
inapllicable for the $L_z-\varphi$ pair?
\item \textbf{Q.3}: Must the set of improvised relations (\ref{eq:6})-
(\ref{eq:14}) be accepted as a natural and justified thing within
the healthful framework of physics ?
\end{itemize}

Related to \textbf{Q.}1 we opine that the mentioned facts are
depreciative elements for TIUR because they increase the deadlook
disclosed by the shortcomings \textbf{Shc.1}-\textbf{4}. The
gravity of the respective deadlook and its  consequences will be
dicussed in Sec.IV.

As regards \textbf{Q.2} the answer is negative. In the next
section we prove efectively that the usual QM procedures are
accurately applicable for $L_z-\varphi$ pair  in respect to all
the physical situations. Based on the respective proof in Sec.IV
we find a credible verdict of non-acceptance for relations
mentioned in \textbf{Q.3}.

\section{The usual QM procedures for $L_z-\varphi$ pair}
The goal of this section is to prove that, in contradiction with a
largely accredited belief, the usual QM procedures are accurately
applicable for the $L_z-\varphi$ pair. The procedures in question
regard the operator-based respectively the Fourier transforms
approaches. Details about the main particularities of the alluded
approaches for the case of $L_z-\varphi$ pair are presented below
in the subsections 3.1 respectively 3.2.

\subsection{The  operator-based approach}
In order to present this approach for the problems of
$L_z-\varphi$ pair let us consider a quantum system in a
rotational motion irrespective of circular or NCR type. The state
of the system is regarded as described by the wave function
$\Psi(q,\varphi)$, indifferently of the fact that $\varphi\in
[0,2\pi]$ (as in (\ref{eq:1}) and (\ref{eq:21})) or
$\varphi\in(-\infty, \infty)$ (as in (\ref{eq:17})). In
$\Psi(q,\varphi)$ by $q$ we denote the orbital coordinates other
than $\varphi$ and specific for the considered system  (e.g. $q$
is: (i) absent in the cases of a motion on a circle or of a
rotator with fixed axis, (ii) the polar angle $\theta$ in the
cases of a particle on a sphere or of a rotator with a mobile
axis, respectively, (iii) the ensamble of both polar angle
$\theta$ and radial distance $r$ in the case of an atomic
electron). In the functions space to which belong
$\Psi(q,\varphi)$ the scalar product is defined as
\begin{equation}
  \label{eq:25}
  (\Psi_1,\Psi_2)=\int \Psi_1^* (q,\varphi)\, \Psi_2 (q,\varphi)\,
  d\Omega_q \, d\varphi
\end{equation}
where $d\Omega_q$ denotes the infinitesimal ``volume'' associated
with the variables $q$ (i.e. $d\Omega_q=\sin\theta \, d\theta$ or
$d\Omega_q=r^2\sin\theta \, dr\, d\theta$ in the above mentioned
cases (ii) respectively (iii)).

For the considered system $L_z$ and $\varphi$ are basic
observables described by the operators $\hat{L}_z$ and
$\hat{\varphi}$ mentioned in (\ref{eq:2}) and (\ref{eq:3}).
Associated with the respective observables we use the following
quantities:
\begin{equation}
  \label{eq:26}
  \langle A \rangle = (\Psi,\hat{A}\Psi)
\end{equation}
\begin{equation}
  \label{eq:27}
  \mathcal{C}(A,B) = (\delta\hat{A}\Psi,\delta\hat{B}\Psi)\; , \qquad
  \delta\hat{A}=\hat{A}-\langle A\rangle
\end{equation}
\begin{equation}
  \label{eq:28}
  \Delta A=\sqrt{\mathcal{C}(A,A)}=(\delta\hat{A}\Psi,\delta\hat{B}
  \Psi)^{1/2}
\end{equation}
which denote: $\langle A\rangle$ = the mean (or expected) value of
the observable $A$, $\mathcal{C}(A,B)$ = the correlation of $A$
and $B$ respectively $\Delta A$ = standard deviation of $A$. Note
that in terms of usual QM the quantities
(\ref{eq:26})-(\ref{eq:28}) are probabilstic parameters
(characteristics): of first order $\langle A\rangle$, respectively
of second order - $\mathcal{C}(A,B)$ and $\Delta A$.

In terms of the above-presented notations the following Schwartz relation
is always satisfied:
\begin{equation}
  \label{eq:29}
  (\delta\hat{L}_z\Psi,\delta\hat{L}_z\Psi)\cdot(\delta\hat{\varphi}\Psi,
  \delta\hat{\varphi}\Psi)\geqslant \vert (\delta\hat{L}_z\Psi,
  \delta\hat{\varphi}\Psi)\vert^2
\end{equation}
which gives directly
\begin{equation}
  \label{eq:30}
  \Delta L_z \cdot \Delta\varphi \geqslant \vert (\delta\hat{L}_z\Psi,
  \delta\hat{\varphi}\Psi)\vert
\end{equation}
This is a general $L_z-\varphi$ relation, valid for all types of rotational
motions.

A particular but restrictive $L_z-\varphi$ relation can be
obtained from (\ref{eq:30}) as follows. If in respect to the
operators $\hat{A}_1=\hat{L}_z$ and $\hat{A}_2$ the wave function
$\Psi$ of the system satisfy the conditions
\begin{equation}
  \label{eq:31}
  (\hat{A}_j \Psi,\hat{A}_k \Psi)=(\Psi,\hat{A}_j \hat{A}_k \Psi) \; ,
  \qquad j=1,2 \; ; k=1,2
\end{equation}
one can write
\begin{eqnarray}
  \label{eq:32}
  (\delta\hat{L}_z\Psi,\delta\hat{\varphi}\Psi)&=\frac{1}{2}\left\langle
  \left[\delta\hat{L}_z,
  \delta\hat{\varphi}\right]_+ \right\rangle + \frac{1}{2}\left\langle
  \big[\hat{L}_z,\hat{\varphi}\big] \right\rangle  \nonumber \\
   &=\frac{1}{2}\left\langle\left[\delta\hat{L}_z,
   \delta\hat{\varphi}\right]_+ \right\rangle-i\,\frac{\hbar}{2}
\end{eqnarray}
where $\langle[\delta\hat{L}_z,\delta\hat{\varphi}]_+\rangle=
\langle\delta\hat{L}_z\delta\hat{\varphi}+\delta\hat{\varphi}
\delta\hat{L}_z\rangle$ is a real quantity. Then by using
(\ref{eq:32}) from (\ref{eq:30}) one obtains the restricted
formula
\begin{equation}
  \label{eq:33}
  \Delta L_z \cdot \Delta\varphi \geqslant \frac{\hbar}{2}
\end{equation}
This is just the disputed and ``prohibited'' relation (\ref{eq:5}).

The above results ensure a veridical base for a natural resolution
of the known disputes regarding the relation
(\ref{eq:33})/(\ref{eq:5}) respectively the applicability of the
usual QM procedures for the observables $L_z$ and $\varphi$. The
mentioned base has to incorporate obligatorily the whole ensamble
of the following evident mathematical findings (\textbf{MF}):
\begin{itemize}
\item \textbf{MF.1}: The general relation (\ref{eq:30}) as well as its
restricted form (\ref{eq:33})/(\ref{eq:5}) are obtainable through
rigurous and precisely specified mathematical ways.
\item \textbf{MF.2}: In the cases described by (\ref{eq:1}) one finds
  \begin{equation}
    \label{eq:34}
    (\hat{L}_z\Psi,\hat{\varphi}\Psi)-(\Psi,\hat{L}_z\hat{\varphi}\Psi)=
    i\hbar
  \end{equation}
and, consequently, the conditions (\ref{eq:31}) are not satisfied.
This means that for such cases the relation
(\ref{eq:33})/(\ref{eq:5}) is not mathematically applicable.
However even in the respective cases the general relation
(\ref{eq:30}) remains valid, degenerating into the trivial
equality $0=0$.
\item \textbf{MF.3}: The conditions (\ref{eq:31}) are always satisfied in the case of
QTP described by (\ref{eq:17}). Consequently for the respective
case the relation (\ref{eq:33})/(\ref{eq:5}) is mathematically
applicable. But note that for the same case the general relation
(\ref{eq:30}) remains also valid.
\item \textbf{MF.4}: For the cases described by (\ref{eq:21}) one obtains:
  \begin{eqnarray}
    \label{eq:35}
    (\hat{L}_z\Psi_l,\hat{\varphi}\Psi_l)-(\Psi_l,\hat{L}_z\hat{\varphi}
    \Psi_l)= \nonumber \\
    =i\hbar \left\{ 1+2\, \mathrm{Im} \left[ \sum_{m=-l}^{l}
    \sum_{m'=-l}^{l} c_m^* c_m m
    \left(Y_{lm},\hat{\varphi} Y_{lm'}\right)\right]\right\}
  \end{eqnarray}
(with $\mathrm{Im}F$ = imaginary part of $F$). Then it results
that for such cases the conditions (\ref{eq:31}) are satisfied or
violated in the situations when the expression after the equality
sign in (\ref{eq:35}) is null respectively non-null.
Corespondingly one finds situations in which the relation
(\ref{eq:33})/(\ref{eq:5}) is true respectively wrong. But note
that in both types of situations the general relation
(\ref{eq:30}) remains valid.
\end{itemize}

Now, in order to facilitate a discussion in Sec.IV, let us refer
to the pair $\theta-\varphi$ (polar and azimuthal angles) considered for the
situation described by (\ref{eq:21}). For such a situation,
similarly with (\ref{eq:30}), it is satisfied the relation
\begin{equation}
  \label{eq:36}
  \Delta\theta \cdot \Delta \varphi \geqslant \left\vert ( \delta
  \hat{\theta}\,\Psi, \delta \hat{\varphi}\,\Psi) \right\vert
\end{equation}
In this relation we take $\hat{\theta}=\theta\cdot$ and the usual
notations introduced above. Additionally we specify that for the
right hand side term from (\ref{eq:36}) we have
\begin{equation}
  \label{eq:37}
  (\delta \hat{\theta}\,\Psi, \delta \hat{\varphi}\,\Psi) =
  \langle \theta\,\varphi\rangle
  -\langle\theta\rangle \cdot \langle\varphi\rangle
\end{equation}
with the detailed notation
\begin{equation}
  \label{eq:38}
  \langle A\rangle =(\Psi,\hat{A}\Psi) = \sum_{m=-l}^{l} \sum_{m'=-l}^{l}
  c_m^* c_{m'} (Y_{lm},\hat{A}Y_{lm'})
\end{equation}
From (\ref{eq:37}) and (\ref{eq:38}) one obseves that, depending
on the concrete values of the coefficients $c_m$, one can find
situations when the quantity $(\delta\hat{\theta}\,\Psi, \delta
\hat{\varphi}\,\Psi)$ have a non-null value. In such situations the
relation (\ref{eq:36}) is satisfied as a formula with a non-null
term in its right hand side.

\subsection{The Fourier transforms approach}

The particularities of $L_z-\varphi$ pair in respect with various
quantum rotational motions can also be approached in terms of
Fourier transform as follows.

Firstly let us refer to the case of QTP described by the wave
functions $\Psi=\Psi(\varphi)$ given by (\ref{eq:17}) and defined
for $\varphi\in(-\infty,\infty)$. For the respective case the
alluded approach was not discussed in literature but it can be
managed by analogy with the known treatment \cite{28} of the
Cartesian coordinate $x$ and momentum $p$ for
$x\in(-\infty,\infty)$. Then with $\Psi(\varphi)$ we associate the
Fourier transform $\tilde{\Psi}(k), k\in(-\infty,\infty)$, defined
by
\begin{equation}
  \label{eq:39}
  \widetilde{\Psi}(k)=\frac{1}{\sqrt{2\pi}} \int_{-\infty}^{\infty}
  \Psi(\varphi) \, e^{-ik\varphi} \, d\varphi
\end{equation}
Due to the Parseval theorem as well as to the fact that
$\Psi(\varphi)$ is normalised we can write
\begin{equation}
  \label{eq:40}
  \int_{-\infty}^{\infty} \vert \Psi(\varphi) \vert^2 d\varphi=
  \int_{-\infty}^{\infty} \vert \widetilde{\Psi}(k) \vert^2 dk = 1
\end{equation}
This relation  shows that both $\vert\Psi(\varphi)\vert^2$ and
$\vert \widetilde{\Psi}(k)\vert^2$ can be regarded as probability
densities for the random variables $\varphi$ respectively $k$. The
mean (or expected) values of the quantities $A=A(\varphi)$ and
$B=B(k)$ depending on the respective variables are defined as
\begin{equation}
  \label{eq:41}
  \langle A\rangle = \int_{-\infty}^{\infty} A(\varphi)\vert\Psi(\varphi)
  \vert^2 d\varphi
\end{equation}
\begin{equation}
  \label{eq:42}
  \langle B\rangle = \int_{-\infty}^{\infty} B(k)\vert\widetilde{\Psi}(k)
  \vert^2 dk
\end{equation}
The following relation is evidently true
\begin{equation}
  \label{eq:43}
  \int_{-\infty}^{\infty} \left\vert \lambda(\varphi-\langle\varphi\rangle)
  \, \Psi(\varphi)+\left(\frac{d}{d\varphi}-i\langle k\rangle\right)
  \Psi(\varphi)\right\vert^2
  d\varphi \geqslant 0
\end{equation}
if $\lambda$ is an arbitrary real parameter. By means of some
simple calculations from (\ref{eq:43}) one obtains
\begin{equation}\label{eq:44}
  \left\langle \left(k-\langle k\rangle\right)^2\right\rangle\cdot
  \left\langle \left(\varphi-\langle \varphi\rangle\right)^2\right\rangle
  \geqslant \frac{1}{4}
\end{equation}
Now by using the notation $\Delta A=\langle (A-\left\langle
A\rangle)^2\right\rangle^{1/2}$ and taking $\hbar k=L_z$ from
(\ref{eq:44}) one finds directly
\begin{equation}\label{eq:45}
  \Delta L_z \cdot \Delta\varphi \geqslant \frac{\hbar}{2}
\end{equation}
i.e. just the relation (\ref{eq:33})/(\ref{eq:5}) which is true in
the considered case of QTP.

In some publications one finds attempts of approaching  with
Fourier transforms the ``periodic situations'' in which $\varphi
\in [0,2\pi]$ and $\Psi(0)=\Psi(2\pi)\neq 0$. The respective
attempts can be resumed as follows. For the corresponding wave
functions $\Psi(\varphi)$ are defined the Fourier coefficients
\begin{equation}\label{eq:46}
  b_m=\frac{1}{\sqrt{2\pi}} \int_{0}^{2\pi} \Psi(\varphi)\,
  e^{-im\varphi} d\varphi
\end{equation}
with $m=0,\pm 1,\pm 2,\ldots$. Based on the fact that
$\Psi(\varphi)$ is normalised on the range $[0,2\pi]$ as well as
on the Parseval theorem  one writes
\begin{equation}\label{eq:47}
  \int_{0}^{2\pi} \vert \Psi(\varphi) \vert^2 d\varphi=
  \sum_m \vert b_m \vert^2 =1
\end{equation}
Then the quantities $\vert \Psi(\varphi) \vert^2$ and $\vert b_m
\vert^2$ are regarded as probability density respectively
probabilities for the continuous respectively discrete random
variables $\varphi$ and $m$. For quantities $A=A(\varphi)$ and
$B=B(m)$ depending on the variables $\varphi$ and $m$ the mean (or
expected) values are
\begin{equation}\label{eq:48}
  \langle A\rangle = \int_0^{2\pi} A(\varphi)\vert \Psi(\varphi)
  \vert^2 d\varphi
\end{equation}
\begin{equation}\label{eq:49}
  \langle B\rangle = \sum_m B(m) \vert b_m \vert^2
\end{equation}
With $\lambda$ as an arbitrary real parameter the evident relation
\begin{equation}\label{eq:50}
  \int_0^{2\pi} \left\vert \lambda (\varphi-\langle\varphi\rangle)
\Psi(\varphi) +\left( \frac{d}{d\varphi} -i\langle m\rangle
\right) \Psi(\varphi) \right\vert^2 d\varphi \geqslant 0
\end{equation}
together with the condition $\Psi(0)=\Psi(2\pi)\neq 0$ give
\begin{equation}\label{eq:51}
  \langle(m-\langle m\rangle)^2\rangle \cdot
  \langle(\varphi-\langle \varphi\rangle)^2\rangle \geqslant
  \frac{1}{4} \left( 1-2\pi \vert \Psi(2\pi)\vert^2\right)^2
\end{equation}
Taking $m\hbar=L_z$ and $\Delta A=\langle(A-\langle
A\rangle)^2\rangle^{1/2}$ from (\ref{eq:51}) results directly
\begin{equation}\label{eq:52}
  \Delta L_z \cdot \Delta\varphi \geqslant \frac{\hbar}{2}
  \left\vert 1-2\pi \vert \Psi(2\pi) \vert^2\right\vert
\end{equation}
So one finds in fact the relation (\ref{eq:15}) which sometimes in
publications is mentioned as $L_z-\varphi$ UR.

Now let us inspect the applicability of the relation (\ref{eq:52})
to the possible ``periodic situations'', when
$\Psi(q,0)=\Psi(q,2\pi)$. Note that the category of such
situations includes the cases of circular motions described by
(\ref{eq:1}) as well as the NCR refered in (\ref{eq:21}). It is to
easy see that in the first cases (\ref{eq:52}) is applicable under
the form of trivial equality $0=0$.

The cases refered to they in (\ref{eq:21}), even if they belong to
the ``periodic situations'' as defined above, they remain outside
the field of applicability for (\ref{eq:52}). This because for the
respective cases the sequence of relations
(\ref{eq:46})-(\ref{eq:52}) as well as the implied quantities must
be adequately modified. So for $\Psi_l(\theta,\varphi)$ given by
(\ref{eq:21}) instead of constant coefficients $b_m$ from
(\ref{eq:46}) it is necessary to operate with the
$\theta$-dependent coefficients
\begin{equation}\label{eq:53}
  b_m(\theta)=\frac{1}{\sqrt{2\pi}} \int_{0}^{2\pi}
  \Psi_l(\theta,\varphi) \, e^{-im\varphi} d\varphi =c_m
  \Theta_{lm} (\theta)
\end{equation}
In (\ref{eq:53}) the coefficients $c_m$ are the same as in
(\ref{eq:21}) while $\Theta_{lm}(\theta)$ denote the $\theta$-
dependent part of the spherical function $Y_{lm}(\theta,\varphi)$
(which can be written as
$Y_{lm}(\theta,\varphi)=\Theta_{lm}(\theta)\,(2\pi)^{-1/2}\,e^{im\varphi}$).
For the here discussed cases instead of (\ref{eq:50}) we have
\begin{eqnarray}\label{eq:54}
\int_{0}^{2\pi} \sin \theta \,d\theta \int_{0}^{2\pi} \vert
\Psi_l(\theta,\varphi)\vert^2 d\varphi= \nonumber \\
=\sum_{m=-l}^{l} \int_{0}^{2\pi} \vert b_m(\theta) \vert^2
\sin\theta \, d\theta =\sum_{m=-l}^{l} \vert c_m\vert^2 =1
\end{eqnarray}
Also for the mean values instead of (\ref{eq:48})-(\ref{eq:49}) we
must take the expressions
\begin{equation}\label{eq:55}
  \langle A\rangle =\int_0^{2\pi} \sin \theta \, d\theta
  \int_0^{2\pi} A(\varphi) \vert \Psi_l(\theta,\varphi)\vert^2
  d\varphi
\end{equation}
\begin{equation}\label{eq:56}
  \langle B\rangle =\sum_{m=-l}^{l} \int_{0}^{\pi} \sin \theta \,
  B(m) \, \vert b_m(\theta)\vert^2 d\theta =\sum_{m=-l}^{l} B(m)
  \vert c_m \vert^2
\end{equation}
For the NCR described by (\ref{eq:21}) instead of (\ref{eq:50}) we
have to operate with the relation
\begin{equation}\label{eq:57}
  \fl \int_0^{2\pi} \sin\theta \, d\theta \int_0^{2\pi}\left\vert
  \lambda (\varphi-\langle \varphi\rangle)\,\Psi_l(\theta,\varphi) +
  \left(\frac{d}{d\varphi} -i\langle m\rangle\right) \Psi_{lm}
  (\theta,\varphi) \right\vert^2 d\varphi \geqslant 0
\end{equation}
From this relation, by taking $\hbar m=L_z$ and through some
simple calculations, one finds the formula
\begin{equation}\label{eq:58}
  \Delta L_z\cdot \Delta\varphi \geqslant \frac{\hbar}{2}\,
  \Big\vert 1-\sum_{m=-l}^{l} \sum_{m'=-l}^{l} c_m^* c_{m'}\,
  \gamma_{mm'}\Big \vert
\end{equation}
where
\begin{equation}\label{eq:59}
  \gamma_{mm'} =\int_0^{2\pi} \Theta_{lm}(\theta)\,
  \Theta_{lm'}(\theta) \sin\theta \, d\theta
\end{equation}
One can conclude that, from the Fourier transforms perspective,
for the NCR described by (\ref{eq:21}) the true $L_z-\varphi$
relation is given by (\ref{eq:58}). In the respective relation the
right hand term can be a null or non-null quantity depending on
the concrete values of the coefficients $c_m$. Note here the fact
that, even for $m\neq m'$ the quantities $\gamma_{mm'}$ (defined
by (\ref{eq:59})), can have non-null values (e.g. when $l=2$ for
$m=1$ and $m'=-1$, respectively for $m=2$ and $m'=-2$ one finds
$\gamma_{mm'}=1$.

In the end of this section we note that the above presented
considerations offer the complete set of elements required for a
correct placement of the $L_z-\varphi$ pair within the
mathematical framework of usual QM procedures.

\section{Concluding remarks}
Now let us discuss the significance of the above findings for the
$L_z-\varphi$ problems. As it was shown the respective problems
regard the situation of observables $L_z$ and $\varphi$ within QM
and related questions of interpretation. The alluded problems
constitute the subject of disputes generated by the unsuccessful
searches to adjust the $L_z-\varphi$ pair to the TIUR assertions.
It is interesting to evaluate the significance of the previous
findings for the mentioned disputes. The respective evaluation
requires firstly an adequate presentation of TIUR. We try to do
such a presentation here below.

The doctrine of TIUR germinates from the preocupation for giving a
unique and generic interpretation for Heisenberg's relations
\begin{equation}\label{eq:60}
  \Delta A \cdot \Delta B \geqslant \frac{1}{2}\,\vert \langle
  [\hat{A},\hat{B}]\rangle\vert
\end{equation}
\begin{equation}\label{eq:61}
  \Delta_{TE} A \cdot \Delta_{TE} B \geqslant \hbar
\end{equation}
Note that, on the one side, the relations (\ref{eq:60}) are QM
formulas written in terms of usual notations  specified in
Sec.III. They were introduced originally \cite{29} by Heisenberg
for the observables $x$ and $p$ (Cartesian coordinate and
momentum). The modern form (\ref{eq:60}), that regard arbitrary
observables $A$ and $B$, were introduced afterwards by other
scientists (see \cite{30,31,32,33,34,35}). On the other side the
relations (\ref{eq:61}) are of ``thought-experimental'' (TE) or
``mental'' nature. They were introduced by means of some
``thought'' or ``mental'' experiments. In relations (\ref{eq:61})
the observables $A$ and $B$ are considered as being canonically
conjugated. Note that the respective relations were introduced
firstly \cite{29} for pairs of observables $x-p$ respectively
$E-t$ (energy-time). Later they were adjusted \cite{36} for the
$L_z-\varphi$ pair by using only some ``mental'' considerations
(i.e. without any real experimental justification).

Starting from (\ref{eq:60}) and (\ref{eq:61}) TIUR was promoted as
a new doctrine (for a bibliography of the significant publications
in the field see \cite{30,31,32,33,34,35,17}. The UR
(\ref{eq:60})-(\ref{eq:61}) have a large popularity, being
frequently regarded as crucial formulas of physics or \cite{35}
even as expression of ``the most important principle of the
twentieth century physics''. But as, as a strange aspect, in the
partisan literature TIUR is often presented so fragmentary and
esoteric that it seems to be rather a dim conception than a
well-delimited scientific doctrine. However, in spite of such an
aspect, from the mentioned literature one can infer that in fact
TIUR  is reducible to the following set of main assertions
(\textbf{Ass.}):
\begin{itemize}
  \item \textbf{Ass.1}: The quantities $\Delta A$ and $\Delta_{TE}A$ from
  (\ref{eq:60}) and (\ref{eq:61}) have  similar significances of measurement
  uncertainty for the QM observable $A$.
  \item \textbf{Ass.2}: With (\ref{eq:60}) and (\ref{eq:61}) the same
  generic interpretation of uncertainty relations (UR) for simultaneous
  measurements regarding the observables $A$ and $B$ it is associated.
  \item \textbf{Ass.3}: For a solitary QM observable $A$ the
  quantity $\Delta A$ can be indefinitely small (even null).
  \item \textbf{Ass.4}: In the case of two ``compatible'' observables
  $A$ and $B$ (when $[\hat{A},\hat{B}]=0$), considered  simultaneously,
  the corresponding quantities $\Delta A$ and $\Delta B$ are mutually
  independent, each of them being allowed to take an idefinitely small
  (even null) value.
  \item \textbf{Ass.5}: For two ``incompatible'' observables $A$ and
  $B$ (when $[\hat{A},\hat{B}]\neq 0$), considered simultaneously,
  the quantities $\Delta A$ and $\Delta B$ are mutually interdependent.
  Their product $\Delta A\cdot\Delta B$ is lower limited by a non-null
  quantity dependent on the Planck's constant $\hbar$.
  \item \textbf{Ass.6}: The relations (\ref{eq:60}) and (\ref{eq:61}) are
  typically QM formulas and they, as well as the Planck's constant $\hbar$,
  do not have analogues in classical (non-quantum) physics.
\end{itemize}

The above assertions constitute the true reference points for the
announced evaluation of the significance of the findings from
Sec.II and Sec.III. Connected with the respective points the
mentioned evaluation is expressible in terms of the following
remarks (\textbf{Rem.}):
\begin{itemize}
  \item \textbf{Rem.1}: First of all we note that the ``mental''
  improvisations (\ref{eq:61}) must not be taken into account in evaluation
  of $L_z-\varphi$ problems or other  questions of interest for physics.
  This because the relations (\ref{eq:61}) have only a provisional character
  while any respectable doctrine (as TIUR claims to be) must operate with
  permanent concepts and relations. The alluded character results from the
  fact that relations (\ref{eq:61}) were founded on old classical limitative
  criteria (introduced by Abbe and Rayleigh - see \cite{37}). But in modern
  experimental physics are known \cite{38,39,40} some super-resolution
  techniques that overstep the respective criteria. This means that
  instead of the quantities $\Delta_{TE}A$ and $\Delta_{TE}B$ from
  (\ref{eq:61}) are imaginable some super-resolution TE (SRTE)
  uncertainties $\Delta_{SRTE}A$ and $\Delta_{SRTE}B$. Then it is possible
  to replace (\ref{eq:61}) with a SRTE relation of the form
\begin{equation}\label{eq:62}
  \Delta_{SRTE}A\cdot\Delta_{SRTE}B < \hbar
\end{equation}
  Such a possibility clearly evidences the provisional and fictitious
  character of the relations (\ref{eq:61}). The respective evidence
  incriminates TIUR in connection with one of its points of origin,
  expressed by \textbf{Ass.1} and \textbf{Ass.2}. Consequently one can
  conclude that in in the discussions of $L_z-\varphi$ problems the relation
  (\ref{eq:61}) must be completely ignored.
  \item \textbf{Rem.2}: From \textbf{Rem.1} it directly results that for
  the debates about the $L_z-\varphi$ problems remains of interest only
  the relation (\ref{eq:60}) and formulas from its  family. Note that
  in fact the alluded problems originate directly from the respective
  relation. On the other hand the formulas from the mentioned family
  operate with quantities  extracted from the mathematical procedures of
  usual QM. But the known huge number of succesful applications confirms
  the correctness of the respective procedures. Particularly this means
  that for the $L_z-\varphi$ pair the results from the approaches above
  notified in Sec.III must be regarded as rigurous and indubitable findings.
  At the actual stage of scientific progress, it is senselessly to contest
  the respective results. Then the natural conclusion is that, in the
  $L_z-\varphi$ case, the debates have to accept the bare (mathematical)
  results of usual QM procedures and to reconsider the (physical or even
  philosophical) interpretation of the respective results.
  \item \textbf{Rem.3}: In the line with the conclusion from \textbf{Rem.2}
  let us refer to the term ``uncertainty'' used by TIUR. The term regards
  the quantities like $\Delta A$ from (\ref{eq:60}) or, similarly the
  quantities $\Delta L_z$ and $\Delta\varphi$ from the relations
  presented in Sec.III. We think that the respective term is
  groundless because of the following facts. Being defined in the
  mathematical QM framework $\Delta A$ signifies a probabilistic parameter
  (standard deviation) of the observable $A$ regarded as a random variable.
  The mentioned framework deals with theoretical concepts and models
  regarding the intrinsic (inner) properties of the considered systems but
  not with elements referring to the (possible) measurements performable
  on the respective systems. Consequently, for a physical system,
  $\Delta A$ refers to the intrinsic characteristics,
  reflected in the fluctuations (deviations from the mean value) of the
  observable $A$. So considered, in spite of the assertions \textbf{Ass.1}
  and \textbf{Ass.2} the quantity $\Delta A$ has no connection with the
  performances (or ``uncertainties'') of the possible measurements
  regarding the observable $A$. Moreover, for a system in a given state,
  $\Delta A$ has a unique and well defined value (connected with the
  corresponding wave function) but not conjectural and changeable
  evaluations (as it is asserted in \textbf{Ass.3}, \textbf{Ass.4} and
  \textbf{Ass.5}).
  \item \textbf{Rem.4}: The above alluded conjectural evaluations can be
  associated with the measurements errors (due to the possible modifications
  in the performances of the measuring devices and techniques). But, as a
  general rule, they regard all the characteristics (i.e. the mean values
  and fluctuations) for every physical observable. Moreover such evaluations
  refer both to the quantum and to the classical physical observables,
  without any essential differences. Also, on a well-principled base,
  one can state that the description of measurements must not pertain
  to QM or to other chapters of theoretical physics. Probably that such
  a statement is not agreed by many of the TIUR partisans. However,
  we opine that the respective statement is consonant with the thinking
  \cite{43}: ``the word `measurement' has been so abused in quantum
  mechanics that it would be good to avoid it altogether''.
  In the spirit of the mentioned statement the QM, as well as the whole
  theoretical physics, must be concerned only with the (conceptual and
  mathematical) models  for the description of the intrinsic properties
  (characteristics) of physical systems.
  \item \textbf{Rem.5}: The quantities $\Delta L_z$ and $\Delta\varphi$
  are rigorously implied in diverse relations, corresponding to various
  rotational motions (see Sec.III). The respective relations are not always
  consonant with the formula (\ref{eq:60}) agreed by TIUR . But such a
  dissonance clearly incriminates the TIUR assertion \textbf{Ass.5}.
  \end{itemize}
  
  The above remarks, directly connected with the $L_z-\varphi$ pair,
  disclose irregularities of the TIUR doctrine. The same irregularities
  as well as additional ones are revealed by some other facts that do not have
  direct connections with the $L_z-\varphi$ pair. We try to present to
  present such facts in the following remarks.
  
  \begin{itemize}
  \item \textbf{Rem.6}: In dispures regarding the applicability of TIUR
  it is also known \cite{9} the case of observables $N-\Phi$
  (number-phase). It is easy to see \cite{23} that, from the perspective
  of usual QM procedures, the case of $N-\Phi$ pair can be approached on a
  way completely analogue with the one used above in subsection 3.1 for
  $L_z-\varphi$ pair. This means that for TIUR the $N-\Phi$ pair entails
  an incrimination similar to the one mentioned in \textbf{Rem.5}.
  \item \textbf{Rem.7}: In Sec.III it was shown that it is possible to exist
  situations in which the angular observables $\theta$ and $\varphi$ satisfy
  the relation (\ref{eq:36}) with a non-null value for the term from the
  right hand side. But as $\theta$ and $\varphi$ are commutable
  ($[\hat{\theta},\hat{\varphi}]=0$) the mentioned situations are in posture
  to incriminate clearly the assertion \textbf{Ass.4} of TIUR. Note that a
  similar situation can also be evidenced \cite{23} for the case of other
  commutable observables (such is the case for the Cartesian coordinates
  $x$ and $y$ regarding a particle in a bi-dimensional potential well
  having inclined walls in respect with $x-y$ axes).
  \item \textbf{Rem.8}: As it is known TIUR promoted the idea  that two
  observables $A$ and $B$ are denotable with the terms ``compatible''
  respectively ``incompatible''  subsequently of the fact that their
  operators $\hat{A}$ and $\hat{B}$ are commutable
  (\mbox{$[\hat{A},\hat{B}]=0$})
  or no ($[A,B]\neq 0$). The mentioned terms are directly connected
  with the fact that the product $\Delta A\cdot \Delta B$
  does not have respectively has a non-null lower limit. Now one can see
  directly that the facts presented in the remarks \textbf{Rem.2,5,6} and
  \textbf{7} proves the desuetude of the respective TIUR idea.
  Particularly the alluded TIUR idea becomes self-contradictory in
  the $L_z-\varphi$ case when the lower limit of the product
  $\Delta L_z\cdot\Delta\varphi$ can be both null and non-null.
  Then, strangely, the same quantities $L_z$ and $\varphi$ appears as
  both ``compatible'' and ``incompatible'' observables.
  \item \textbf{Rem.9}: The quantities $\Delta L_z$ and $\Delta\varphi$
  from (\ref{eq:30}) (and similarly $\Delta A$ and $\Delta B$ from
  (\ref{eq:60})) are second order probabilistic parameters.
  Consequently (\ref{eq:30}) is a simple second order probabilistic formula.
  But (\ref{eq:30}) is generalisable in form of some extended relations
  referring also to the second order probabilistic parameters. So one
  obtains \cite{21,22,23,26,16}: (i) bi-temporal relations, (ii) many-observable
  relations and (iii) macroscopic quantum statistical relations.
  For the mentioned extended relations TIUR has to give an interpretation
  concordant with its own essence, if it is a well-grounded conception.
  But to find such an interpretation on natural ways (i.e. without esoteric
  and extra-physical considerations) seems to be a difficult (even imposible)
  task. In this sense it is significant to remind the lack of success
  connected with the above noted macroscopic relations. In order to adjust
  the respective relations to the TIUR assertions, among other things,
  it vas promoted  the idea of an  appeal to the so-called ``macroscopic
  operators'' (see \cite{44} and references). But in fact \cite{23,26}
  the mentioned appeal does not ensure for TIUR  the avoidance of the
  involved shortcomings. Moreover the respective ``macroscopic operators''
  appear as being only fictitious concepts without any real applicability in
  physics. It is also interesting to observe that, in the last years
  the problem of the macroscopic relations and operators is eschewed in the
  partisan literature of TIUR, even if the respective problem still remains
  unclarified from the TIUR perspective.
  \item \textbf{Rem.10}: In classical physics of probabilistic structure
  a nontrivial interest can also present the higher order parameters
  (correlations) (see \cite{45,46}). This fact suggests that in the case
  of QM, additionally to the second order quantities like $\Delta A=
  (\delta\hat{A}\Psi,\delta\hat{A}\Psi)^{1/2}$ or
  $(\delta\hat{A}\Psi,\delta\hat{B}\Psi)$ from (\ref{eq:60}) or (\ref{eq:30}),
  to use also higher order correlations such as
  $((\delta\hat{A})^r \Psi,(\delta\hat{B})^s \Psi)$ with $r+s \geqslant
  3$. Then, naturally, for the respective correlations TIUR has to give
  an interpretation concordant with his own doctrine. But, in our opinion,
  it is less probable (or even excluded) that such an interpretation will be
  promoted by the TIUR partisans.
  \item \textbf{Rem.11}: In contradiction with the assertion \textbf{Ass.6}
  of TIUR in classical (non-quantum) physics \cite{23,47,48} there are really
  some formulas that are completely similar to the QM relations (\ref{eq:60})
  and (\ref{eq:30}). The alluded formulas can be written in the form
\begin{equation}\label{eq:63}
  \Delta_{CF}A\cdot \Delta_{CF}B \geqslant \vert \langle \delta A
  \cdot \delta B\rangle_{CF} \vert
\end{equation}
  where the standard deviations $\Delta_{CF}A$ and $\Delta_{CF}B$
  respectively the correlation $\langle \delta A \cdot \delta B\rangle_{CF}$
  are refering to the classical fluctuations (CF) of the macroscopic
  observables $A$ and $B$. One can see that in fact both classical formulas
  (\ref{eq:63}) and QM relations (\ref{eq:60}) and (\ref{eq:30}) imply only
  second order probabilistic parameters (standard deviations and correlations).
  The respective parameters describe the fluctuations of the corresponding
  observables considered as random variables. Note that the fluctuations
  regard the intrinsic (own) properties of the physical systems but not the
  aspects (uncertainties) of the measurements performed on such systems.
  \item \textbf{Rem.12}: The concrete expressions of the fluctuations
  parameters implied in the quantum-classical similarity mentioned in
  \textbf{Rem.11} evidence the fact \cite{49,48} that the Planck's constant
  $\hbar$ has also a classical similar, namely the Boltzmann's constant $k$.
  It was shown \cite{49} that both $\hbar$ and $k$ play similar roles of
  generic indicators of stochasticity (randomness) in the cases of quantum
  respectively classical observables. The mentioned  roles are directly
  connected with the probabilistic parameters implied in the classical
  formulas (\ref{eq:63}) respectively in the QM  relations (\ref{eq:60}) and
  (\ref{eq:30}). Such parameters are expressible \cite{49} in terms of
  products between $k$ respectively $\hbar$ and quantities which do not
  contain $k$ respectively $\hbar$. But the alluded parameters disclose
  the level of stochasticity (randomness) of the referred observables and
  systems. Then it results that $k$ and $\hbar$ have the attributes of
  stochasticity  (randomness) indicators . The noted attributes are
  generic i.e. they are specific for all the observables of a
  system respectively for all systems from the same class
  (classical or quantum). Note that the above mentioned similarity
  between $\hbar$ and $k$ clearly contradicts the assertion
  \textbf{Ass.6} of TIUR.
  \item \textbf{Rem.13} The discussions included in the above remarks
  \textbf{Rem.1}-\textbf{12} collect new supporting elements for the opinion
  that \cite{50} the uncertainty relations \emph{``are  probably the most
  controverted formulae in the whole of the theoretical physics''}.
  Also through the respective dicussions we update and consolidate the
  convincingness of the observation \cite{51}: \emph{``the idea that there are
  defects in the foundations of orthodox quantum theory is unquestionable
  present in the conscience of many physicists''}.
  \end{itemize}
  
  Now  it is evident that in their whole ensamble the remarks
  \textbf{Rem.1}-\textbf{13} indubitably argue for the following concluding
  remark:
  
  \begin{itemize}
  \item \textbf{Rem.14}: From the perspective of physics  TIUR  must be
  denied as an incorrect and useless doctrine which actually generates
  only senseless and unproductive disputes. Consequently the relations
  (\ref{eq:61}) are  rejected as fictitious formulas while the relations
  (\ref{eq:60}) remain  as simple formulas , deprived of any capital
  (or extraordinary) significance for physics. Then the QM observables
  appear as (generalised) random variables endowed with fluctuations
  characterised by some  ordinary parameters as the standard deviations
  and correlations implied in relations of (\ref{eq:30}) or (\ref{eq:60})
  type. Moreover in the respective relations the commutatibility of the
  corresponding operators do not play any capital (or extaordinary) role.
  This fact associated with \textbf{Rem.8} shows that in connection with
  QM observables the terms ``compatible'' repectively ``incompatible''
  are completely obsolete and useless.
  \end{itemize}
  
  Then, in respect with the $L_z-\varphi$ problems that motivate the
  present work, we are justified to note the next two remarks:
  
  \begin{itemize}
  \item \textbf{Rem.15}: For a correct physical description of the
  observables $L_z$ and $\varphi$ the usual QM procedures (as presented
  in Sec.III) are sufficient and they have not to be adjusted with any
  other esoteric considerations or improvised relations.
  \item \textbf{Rem.16}: The improvised $L_z-\varphi$ UR (\ref{eq:6})-(\ref{eq:14})
  must be rejected as formulas without any authentic physical significance.
  Their associate shortcomings \textbf{Shc.1}-\textbf{4}, noted in Sec.II,
  are completely unsurmountable because they cannot be avoided by means of
  some credible physical arguments.
  \end{itemize}

  In the context of the above noted remarks it is the place to add some
  observations about another question often mentioned in connection with
  TIUR. The question regards the problem of the pair $E-t$ (energy and time).
  We include the announced observations in the next remark.
  
  \begin{itemize}
  \item \textbf{Rem.17}: The $E-t$ problem was largely disputed during the history of QM
  (see \cite{17} and references). In the main the alluded disputes were
  focused on the subordination of the QM description for the $E-t$ pair to
  the relations (\ref{eq:60}) and (\ref{eq:61}). As the respective
  subordonation happened to be neither evident nor a direct one
  there was promoted diverse adjustements (some of them more or
  less exotic). Here we do not intend to examine in details the
  mentioned historical facts or the beliefs germinated from them.
  Our present goal is that, in connection with the E-t problem, to
  note some opinions consonant with the above argued views and
  confluent discussions. The respective opinions are:

  (i) As in fact the relations (\ref{eq:60}) and (\ref{eq:61}) are not
  capital or extraordinary scientific elements the QM description of the
  $E-t$ pair needn't be subordinated to the respective relations.

  (ii) Due to the things presented in \textbf{Rem.1}, similarly to the
  $L_z-\varphi$ case, in the dicussions  about the $E-t$ pair the relation
  (\ref{eq:61}) must be also completely ignored.

  (iii) Connected with the possible significance of the relation (\ref{eq:60})
  it is important to note the following aspects. In many known texts for the
  QM description of the $E-t$ pair one uses the operators
  $\hat{E}=i\hbar \frac{\partial}{\partial t}$ and $\hat{t}=t\cdot$.
  But then it must be noted the truth that related to the respective usance
  the relation (\ref{eq:60}) is inadequate. The inadequacy is due to the
  infringement of the conditions of (\ref{eq:31})-type. Indeed the alluded
  operators are associated with the result
\begin{equation}\label{eq:64}
  (\hat{E}\Psi,\hat{t}\Psi)-(\Psi,\hat{E}\hat{t}\,\Psi)=-i\hbar
\end{equation}
  valid for every wave function $\Psi$. Here the notations are the ones
  from usual QM reminded in Sec.III. The result (\ref{eq:64}) shows that the
  conditions of (\ref{eq:31}) type are not satisfied  and, consequently,
  it proves that (\ref{eq:60}) is not applicable. However, as it was noted
  in Sec.III, independently of a result like (\ref{eq:64}), the relations of
  (\ref{eq:30})-type are true. This means that  with the mentioned operators
  one obtains the relation 
\begin{equation}\label{eq:65}
  \Delta E \cdot \Delta t \geqslant \vert (\delta \hat{E}\Psi,
  \delta\hat{t} \Psi)\vert
\end{equation}
  which reduces itself to the trivial equality $0=0$ because
  $\langle t\rangle =t$ and $\delta\hat{t}=0$. Such an equality has as sole
  physical significance the fact that in the framework of usual QM the time
  $t$ is a deterministic (dispersion free) variable but not a probabilistic
  (random) quantity. Note that in the mentioned framework, essentially
  implied in (\ref{eq:64}) and (\ref{eq:65}), the probabilistic load is
  carried by other variables (of orbital or spin nature) but not by the time.

  (iv) We opine that naturally in QM framework the time $t$ must be regarded
  as a deterministic (non-random) variable, which has no fluctuation
  characteristics (describable in terms of standard deviations and
  correlations). Consequently in the respective framework the time $t$
  needn't be associated with an operator. The so-called ``time operators''
  introduced in some publications can be regarded rather as pieces of pure
  mathematical reasonings. On the other hand we consider that, in the same
  framework, an operator must describe the energy. In a natural conception
  the respective operator is identifiable with the QM Hamiltonian.

  (v) In the disputes about the $E-t$ pair, beside the relation (\ref{eq:60}),
  many publications put forward the ``spreading'' formula
\begin{equation}\label{eq:66}
  \sqcap E \cdot \sqcap t \geqslant \frac {\hbar}{2}
\end{equation}
  specific for the QM wave packets. In association it is promoted the belief
  that, likewise with (\ref{eq:60}), the formula (\ref{eq:66}) is a capital
  indicator for the QM peculiarity (i.e. for the distinction in comparasion
  with classical physics). We think that the respective formula is not an
  indicator of the mentioned kind. In fact it is completely analogous with
  the classical ``spreading'' formula \cite{52}
\begin{equation}\label{eq:67}
  \sqcap\omega \cdot \sqcap t \geqslant \frac{1}{2}
\end{equation}
specific for signal pulses (packets) ($\omega$ = angular frequency
and $t$ = time). Mathematically both (\ref{eq:66}) and
(\ref{eq:67}) can be introduced similarly through relations like
(\ref{eq:44}) from Fourier transforms approach. As significance
in both cases (\ref{eq:66}) and  (\ref{eq:67} $\sqcap t$ denotes
the ``duration'' of the packet (pulse). In classical context
\cite{52} $\sqcap\omega$ signifies the spectral width (or
``spreading'') of the pulses. Similarly for QM wave packets
$\sqcap E$ must be interpreted also as spectal (energetic) width.
As regards the likeness between the relations (\ref{eq:66}) and
(\ref{eq:60}) it is clear only of same nature. This because the
respective relations are introduced within different mathematical
and conceptual contexts.
\end{itemize}

Our above discussions are focused on the remarks about the
physical significance of some relations from the same family with
(\ref{eq:60}). But,  besides  the mentioned significance, the
respective relations seem to present a mathematical attraction
(see \cite{16,53,54} and references). Connected with such a fact
we note the next remark.

\begin{itemize}
\item \textbf{Rem.18}: Even if the relation (\ref{eq:60}) must be regarded as a
formula without any capital or extraordinary physical significance
from a mathematical view point it appears in posture of an
interesting source of inspiration. Such a posture explains the
appreciable number of mathematical formulas which extend or
generalize the relation (\ref{eq:60}) (for a relevant bibliography
in ths sense see \cite{16,53,54} and references).  In the next
years probably the alluded number will increase as a result of
some pure mathematical approaches. The physical significance
(most probably of non-capital importance) of some of the
mentioned mathematical formulas seems to be actually in the
attention of scientists. On the other hand , in connection with
the respective formulas it is of nontrivial interest to take also
into account the idea according to which an increased number of
mathematical formulas and reasonings does not provide without
fail some significant results for physics.
\end{itemize}

Now let us note a few observations on the question of quantum
measurements, mainly generated by TIUR history. According to its
assertions \textbf{Ass.1} and \textbf{Ass.2} TIUR promoted the
idea that the description of the respective measurements must be
associated with the relations (\ref{eq:60}). But, as we have shown
above the respective relations have to be deprived of the
traditionally assumed significance and TIUR must be denied. Then
it results that the alluded question of quantum measurements
remains an unelucidated problem that requires some additional
considerations.  We think that , in a first approximation, the
respective considerations can be stated with the following remark.

\begin{itemize}
\item \textbf{Rem.19}: The descriptions of quantum measurements have to be
done in specific approaches and frameworks that must be distinct and
additional in respect with the usual QM.
\end{itemize}

A possible approach of the mentioned kind was formulated in our
recent work \cite{55}. The basic idea promoted by us is that, for
a quantum microparticle in an orbital motion, the description of
measurements can be done in terms of linear transforms for both
probability density  and probability current. One of the main
advantages of our approach is that it avoids any considerations
connected with the strange idea  about the collapse (reduction)
of wave function.

\ack Along the years, in  the investigations  of here approached
problems, I have studied a large number of publications.  Some of
them  were obtained directly from their authors to whom I express
my deep gratitude. I wish to express my profound thanks to the
World Scientific Company, Singapore, for putting at my disposal a
copy of the monumental book \cite{17}.

Because of various reasons in the present work I confined myself
to a restricted number of references. Doing so I cannot give (and
I did not intend to do) an exhaustive presentation and
appreciation of the huge number of publications in the field.
But, I think, that the mentioned references are works that can
offer good and credible guide marks for the discussions
approached here.

I note here that parts of the investigations reported in the
present text benefited from some  facilities from grants
supported by Roumanian Ministry of Education and Research.

\section*{Appendix: A reply addendum}
Our first opinions about the $L_z-\varphi$ pair were presented in
earlier works \cite{18,19}. The presentations were more modest and
less complete - e.g. we did not use all the  arguments resulting
from the above discussed cases which are described by the wave
functions (\ref{eq:17}) and (\ref{eq:21}). Newertheles, we think
that the alluded opinions were correct in their essence. However
in a review \cite{56} of Prof.F.E.Schroeck the respective our
opinions were judged as being erroneus. In this addendum we wish
to reply to the mentioned judgements by using arguments based on
the considerations included in the present work.

The main error reproached to us in \cite{56} is: \emph{``most of the
results stated concerning angular momentum and angle operators
(including the canonical commutation relations) are false, this
being a consequence of not using  Rieman-Stieljes integration
theory which is necessitated since the angle function has a jump
discontinuity''}.

In order to give an answer to the respective reproach we appeal
to  the following arguments :

(i) One can see that, mainly, the reproach is founded on the
supposition that the variable $\varphi$ has jump (of magnitude
$2\pi$ at $\varphi=0$ or, equivalently, at $\varphi=2\pi$). Such a
supposition implies directly the idea  that for the operators
$\hat{L}_z$ and $\hat{\varphi}$ the commutation relation is not
the one given in (\ref{eq:3}) but
$[\hat{L}_z,\hat{\varphi}]=-i\hbar+i\hbar2\pi\delta$ (where
$\delta$ denotes the Dirac function at the boundary $\varphi=0$ or
$\varphi=2\pi$). We inform that the respective idea was confesed
to us by Prof. F.E. Schroeck in two letters (dated September
16,1981 respectively April2,1982). Note also the fact that, in a
visible or masked manner, the same idea is also promoted in a
number of publications dealing with the $L_z-\varphi$ problem.

(ii)  The reproach could  refer only to the situations described
by  the wave functions (\ref{eq:1}) when  the real physical range
of $\varphi$ is the interval $[0,2\pi]$ (at the respective date
our works \cite{20,22,23} regarding QTP had not appeared yet). But
then the mentioned supposition and idea reveal themselves as
strange things because they imply the hypothesis that the range
of $\varphi$ is larger than the interval $[0,2\pi]$, eventually
even the infinite interval $(-\infty,\infty)$. But it is evident
that such a hypothesis is wrong and that for the situations under
discussions the correct range for $\varphi$ is interval $[0,2\pi]$
It is also important that in the mentioned situations the
normalization of the wave functions is done in fact on the
interval $[0,2\pi]$ but not on other larger domains . Therefore
related to the respective situations it is senseless to consider
that ``the angle function has a jump discontinuity''. Consequently
in the alluded context to ``using  Rieman-Stieljes integration
theory'' is not necessitated at all.

(iii) In the case of QTP deescribed by (\ref{eq:17}) the relation
$[\hat{L}_z,\hat{\varphi}]=-i\hbar$, given in (\ref{eq:3}) is
indubitably applicable. Then , in the spirit of the mentioned
reproach , for the same pair of obsevables $L_z$ and $\varphi$,
one has to tolerate two commpletely dissimilar and irreconcilable
commutation relations $[\hat{L}_z,\hat{\varphi}]=-i\hbar$ and
$[\hat{L}_z,\hat{\varphi}]=-i\hbar+i\hbar 2\pi \delta$. But such a
tolerance is evidently a senseless thing, without any real
(physical) support.

(iv) From the considerations presented above in Sec.III it
results directly the irrebutable conclusion  that, from a
mathematical perspective, we have the unique commutation relation
$[\hat{L}_z,\hat{\varphi}]=-i\hbar$ for all the rotational
motions (described by any of the wave functions (\ref{eq:1}),
(\ref{eq:17}) or (\ref{eq:21})).

(v) The mathematical findings \textbf{MF.1}-\textbf{4} from
Sec.III reveal the fact that, in the descriptions of various
rotational motions, the differences are evidenced not by the
commutation relation for $\hat{L}_z$ and $\hat\varphi$ but by the
resulting formula for $\Delta L_z$ and $\Delta\varphi$. The
respective formula is always valid in the general version
(\ref{eq:30}) but,  depending on the fulfilment of the conditions
(\ref{eq:31}), it can take the particular and restricted form
(\ref{eq:33})/(\ref{eq:5}).

The ensemble of the above noted arguments (i)-(v) proves as
unfounded the reproaches \cite{56} of Prof. F.E. Schroeck
regarding our opinions about the $L_z-\varphi$ pair.

\section*{List of abbreviations}
\begin{tabular}{ll}
\textbf{Ass.} & = assertion \\
CF & = classical fluctuations \\
\textbf{MF} & = mathematical finding  \\
NCR & = non-circular rotations  \\
Q & = question  \\
QM & = quantum mechanics  \\
\textbf{Rem}  & = remark  \\
\textbf{Shc} & = shortcoming  \\
TIUR  & = traditional interpretation of uncertainty relations  \\
UR & = uncertainty relations
\end{tabular}

\end{document}